\documentclass{article}
\usepackage{url}

\usepackage{amssymb}
\usepackage{amsmath}
\usepackage{latexsym}
\def\R{\mathbb{R}}

\def\e{{\epsilon}}
\def\a{{\alpha}}

\newcommand{\old}[1]{{}}

{\makeatletter
 \gdef\xxxmark{%
   \expandafter\ifx\csname @captype\endcsname\relax 
     \marginpar{xxx}
   \else
     xxx 
   \fi}
 \gdef\xxx{\@ifnextchar[\xxx@lab\xxx@nolab}
 \long\gdef\xxx@lab[#1]#2{{\bf [\xxxmark #2 ---{\sc #1}]}}
 \long\gdef\xxx@nolab#1{{\bf [\xxxmark #1]}}
 \long\gdef\xxx@lab[#1]#2{}\long\gdef\xxx@nolab#1{}%
}

\title{Computational Geometry Column 42}
\author{%
Joseph S.~B.~Mitchell\thanks{
Dept. of Applied Mathematics and Statistics,
University at Stony Brook,
Stony Brook, New York 11794-3600, USA.
jsbm@\allowbreak ams.\allowbreak sunysb.\allowbreak edu.
Supported by HRL Laboratories,
NSF Grant CCR-9732221, NASA Ames Research
Center, Northrop-Grumman Corporation, Sandia National Labs, 
and Sun Microsystems.
}
\and
Joseph O'Rourke\thanks{
Dept. of Computer Science, Smith Col\-lege, North\-ampton, 
MA 01063, USA.
orourke@\allowbreak cs.\allowbreak smith.\allowbreak edu.
Supported by NSF Grant CCR-9731804.
}
}
\date{}
\begin{document}
\maketitle

\begin{abstract}
A compendium of thirty previously published open
problems in computational geometry is presented.

\noindent
[\emph{SIGACT News}, {\bf 32}(3) Issue, 120 Sep. 2001, 63--72.]
\end{abstract}

The computational geometry community has made many 
advances in the relatively short (quarter-century) of the
field's existence.
Along the way researchers have 
engaged with a number of problems
that have resisted solution.
We gather here a list of open problems in computational geometry
(and closely related disciplines)
which together have occupied a sizable portion of
the community's efforts over the last decade or more.
We make no claim to comprehensiveness,
only that were all these problems to be solved, the
field would be greatly advanced.
All the problems have appeared in earlier publications,
but we believe all remain open as stated.
We present them in condensed form, without always defining
every technical term, but in each case providing at least one
reference for further investigation.
Our list consists of predominantly theoretical
questions for which the problem can be succinctly stated
and the measure of success is clear.  We do not attempt
here to list the wealth of important problems
in applied and experimental computational
geometry now being addressed by the community as it
responds to the application-driven need for practical geometric algorithms;
we hope that an ongoing project to compile a more
comprehensive list will address this omission.
We encourage correspondence to correct, extend, 
and update a Web version
of this list.\footnote{
        \url{http://cs.smith.edu/~orourke/TOPP/}.
}

\begin{enumerate}

\item Can a \emph{minimum weight triangulation} of a planar point
set---one minimizing the total edge length---be found in polynomial
time?  This problem is 
one
of the few from Garey and Johnson~\cite{gj-cigtn-79} whose
complexity status remains unknown.
The best approximation algorithms achieve a (large) constant times
the optimal length~\cite{lk-qgtam-96};
good heuristics are known~\cite{dmm-naefm-95}.
If Steiner points are allowed, again constant-factor
approximations are known \cite{e-amwst-94,cl-qrsam-98},
but it is open to decide even if a minimum-weight Steiner triangulation
exists (the minimum might be approached only
in the limit).

\old{
"Quadtree, Ray Shooting and Approximate Minimum
Weight Steiner Triangulation", by Siu-Wing Cheng and Kam-Hing Lee
journal version submitted to CGTA  (previous version appears in ISAAC'98)
}

\item What is the maximum number of combinatorial changes
possible in a Euclidean Voronoi diagram of a set of $n$ points each moving 
along a line at unit speed in two dimensions?
The best lower bound known is quadratic, and the best
upper bound is cubic~\cite[p.~297]{sa-dsstg-95}.
If the speeds are allowed to differ, the upper bound
remains essentially cubic~\cite{agmr-vdmp-98}.
The general belief is that the complexity should be
close to quadratic; Chew~\cite{c-nqblv-97} showed this to be the case
if the underlying metric is $L_1$ (or $L_\infty$).

\item What is the combinatorial complexity of the Voronoi diagram
of a set of lines (or line segments) in three dimensions?
This problem is closely related to the previous problem,
because points moving in the plane with constant velocity yield
straight-line trajectories in space-time.  Again, there is a gap between
a lower bound of $\Omega(n^2)$ and an upper bound that is
essentially cubic~\cite{s-atubl-94} for the Euclidean case
(and yet is
quadratic for polyhedral metrics~\cite{bsty-vdhdc-98}).  
A recent advance shows that 
the ``level sets'' of the Voronoi diagram of lines, given by the union of
a set of cylinders, 
indeed has near-quadratic complexity~\cite{as-pckum-00}.

\item What is the complexity of the union of ``fat'' objects
in $\R^3$?  
The Minkowski sum of polyhedra of $n$ vertices
has complexity $O(n^{2+\e})$~\cite{as-pckum-99},
as does
the union of $n$ congruent cubes~\cite{pss-ucc3d-01}.
It is widely believed the same should hold true
for \emph{fat} objects, those with a bounded ratio of
circumradius to inradius, as in does in $\R^2$~\cite{es-cufop-00}.

\item Can the Euclidean
\emph{minimum spanning tree} (MST) of $n$ points in $\R^d$
be computed in time close to the lower bound of
$\Omega( n \log n)$ \cite{gkms-lbrad-96}?
An MST of a connected graph can be computed in time nearly
linear in the number of edges~\cite{c-fdams-97},
but this is quadratic in the number of vertices $n$
for geometric graphs.
And indeed the best upper bounds for the Euclidean
MST approach quadratic for large $d$, e.g.,~\cite{ck-dmpsa-95}.
This problem is intimately related to the bichromatic closest
pair problem~\cite{aesw-emstb-91}.

\xxx{See \cite{aesw-emstb-91}. See also Eppstein~\cite{e-sts-98}.}

\item
What is the complexity of computing a minimum-cost
Euclidean matching for $2n$ points in the plane?
An algorithm that achieves the minimum and runs
in nearly $O(n^{2.5})$ time has long been available~\cite{v-ghm-89}.
Recently Arora showed how to achieve a $(1+\e)$-approximation
in $n (\log n)^{O(1/\e)}$ time~\cite{a-ptase-98}.

\item
What is the maximum number of $k$-sets? (Equivalently, what
is the maximum complexity of a $k$-level in an arrangement of
hyperplanes?)
For a given set $P$ of $n$ points, $S\subset P$ is a {\em $k$-set}
if $|S|=k$ and $S=P\cap H$ for some open halfspace $H$.
Even for points in two dimensions the problem
remains open: The maximum number of $k$-sets as a function of $n$ and $k$
is known to be $O(n k^{1/3})$ by a recent advance of Dey~\cite{d-ibpks-98}, 
while the best lower bound
is only slightly superlinear~\cite{t-psmks-00}.

\item
Is linear programming strongly polynomial?
It is known to be weakly polynomial, exponential in
the bit complexity of the input data~\cite{k-palp-80,k-nptal-84a}.
Subexponential time is achievable 
via a randomized algorithm~\cite{msw-sblp-96}.
In any fixed dimension, linear programming can be solved
in strongly polynomial linear time 
(linear in the input size)~\cite{d-ltatt-84,m-lpltw-84}.

\item
Can every convex polyhedron be cut along its edges and unfolded
flat to a single, nonoverlapping, simple polygon?
The answer is known to be {\sc no} 
for nonconvex polyhedra~\cite{bdems-uptf-01},
but has been long open for convex polyhedra~\cite{s-cpcn-75}\cite{o-fucg-00}.

\item
Is there a deterministic, linear-time polygon triangulation
algorithm significantly simpler than that of Chazelle~\cite{c-tsplt-91a}?
Simple randomized algorithms that are close to linear-time
are known~\cite{s-sfira-91}, 
and a recent randomized linear-time 
algorithm~\cite{agr-lttsp-00} avoids much of the intricacies
of Chazelle's algorithm.  
Relatedly, is there a simple linear-time algorithm for
computing a shortest path in a simple polygon, without
first applying a more complicated triangulation algorithm?

\item Can the class of \emph{$3$-SUM hard} problems~\cite{go-copcg-95}
be solved in subquadratic time?
These problems can be reduced from the problem
of determining whether,
given three sets of integers, $A$, $B$, and $C$ with
total size $n$, there are elements
$a \in A$, $b \in B$, and $c \in C$ such that $a+b=c$.
Many fundamental geometric problems fall in this class;
e.g., computing the area of the union of $n$ triangles.

\item
Can a planar convex hull be maintained to support
both dynamic updates and queries in logarithmic time?
Recently 
the $O(\log^2 n)$ barrier was
broken with a $O( \log^{1+\e} n)$ update- and $O(\log n)$ query-time
structure~\cite{c-dpcho-99},
and the update time further improved to $O(\log n \log\log n)$
in~\cite{bj-dpcho-00}.

\item 
Is there an $O(n)$-space data structure that supports
$O(\log n)$-time point-location queries
in a three-dimensional subdivision of $n$ faces?
Currently $O(n \log n)$ space and $O(\log^2 n)$ queries
are achievable~\cite{s-pl-97}.

\item
Is it possible to construct a \emph{binary space partition} (BSP)
for $n$ disjoint line segments in the plane of size
less than $\Theta(n \log n)$?  The upper bound of $O(n\log n)$
was established by Paterson and Yao~\cite{py-ebsph-90}.  
Recently T\'{o}th~\cite{t-nbpp-01} improved the 
trivial $\Omega(n)$
lower bound to
$\Omega(n\log n/\log\log n)$.  Can the remaining gap be
closed?

\item
What is the best output-sensitive convex hull algorithm
for $n$ points in $\R^d$?
The lower bound is $\Omega( n \log f + f )$ for $f$ facets
(the output size).
The best upper bound
to date is $O( n \log f + (nf)^{1-\delta} \log^{O(1)} n )$
with $\delta = 1/( \lfloor d/2 \rfloor + 1)$~\cite{c-osrch-96},
which is optimal for sufficiently small $f$.

\item Can the number of simple polygonalizations of
a set of $n$ points in the plane be computed in polynomial time?
Certain special cases are known (e.g., 
monotone simple polygonalizations~\cite{zssm-grpgv-96}), 
but the general
problem remains open.  The problem is closely related to
that of generating a ``random'' instance of a simple polygon
on a given set of vertices. 
Heuristic methods are known and implemented~\cite{ah-hgrp-96}.

\item
Given a visibility graph $G$ and a Hamiltonian circuit $C$,
determine in polynomial time whether there is a simple polygon whose
vertex visibility graph is $G$, and whose boundary corresponds to $C$.
The problem is not even known to be in NP~\cite{o-cgc18-93},
although it is
for ``pseudo-polygon'' visibility
graphs~\cite{os-vepvgcr-97}.

\item
When a collection of
disks are pushed closer together, so that no distance
between two center points increases, can the area of
their union increase?  It seems the answer is {\sc no},
but this has only been settled in the continuous-motion case
\cite{bs-pdtcm-98}.
The corresponding question for intersection area decrease
is similarly unresolved~\cite{c-aidp-96}.

\item
What is the complexity of the \emph{vertical decomposition}
of $n$ surfaces in $\R^d$, $d \ge 5$?
The lower bound of $\Omega(n^d)$ was nearly
achieved up to $d=3$~\cite[p.~271]{as-dssga-00},
but a gap remained for $d \ge 4$.
A recent result~\cite{k-atubv-01}
covers $d=4$ and achieves $O(n^{2d-4+\e})$ for general $d$,
leaving a gap for $d \ge 5$.
\xxx[JSBM]{I think that Vladlen Koltun's new FOCS'01
paper implies an upper bound of $O(n^{2d-4})$ now... check!
His bound is tight for 4D and should be cited.}

\item
What is the complexity of computing 
a spanning tree of a planar point set having minimum stabbing number?
The \emph{stabbing number} of a tree $T$ is the maximum number of
edges of $T$ intersected by a line.
Any set of $n$ points in the plane has a
spanning tree of stabbing number $O(\sqrt{n})$~\cite{a-rsoas-92,c-tbsns-88,w-ggssn-93},
and this bound is tight in the worst case.
However, nothing is known about the complexity of computing a spanning tree (or triangulation)
of minimum stabbing number, exactly or approximately.

\item Can shortest paths among $n$ obstacles in the plane,
with a total of $n$ vertices, be found in
optimal $O(n + h \log h)$ time using $O(n)$ space?
The only algorithm that is linear in $n$ in time and space is
quadratic in $h$~\cite{kmm-eaesp-97}; 
$O(n\log n)$ time, using $O(n\log n)$ space, 
is known~\cite{hs-oaesp-99}.
In three dimensions, the Euclidean shortest path problem among general
obstacles is NP-hard, but its complexity remains open
for some special cases, such as when the obstacles are disjoint unit
spheres or axis-aligned boxes; see \cite{m-gspno-00}. 

\item
Can a minimum-link path among polygonal obstacles
be found in subquadratic time?
The best algorithm known requires essentially
quadratic time in 
the worst case~\cite{mrw-mlpop-92}.  What is the complexity of 
computing minimum-link paths in three dimensions?

\item How many $\pi$-floodlights
are always sufficient
to illuminate any polygon of $n$ vertices,
with at most one floodlight placed at each vertex?
An \emph{$\a$-floodlight} is a light of aperture $\a$.
It was established in~\cite{eoux-ipvf-95} that
for any $\a < \pi$, there is a polygon that cannot be
illuminated with an $\a$-floodlight at each vertex.
When $\a=\pi$, $n-2$ lights (trivially) suffice.
So it is of interest 
(as noted in~\cite{u-agip-00})
to learn whether a fraction of
$n$ lights suffice for $\pi$-floodlights.

\item
Can an $n$-vertex polygonal curve be simplified 
in time nearly linear in $n$?  
The goal is to compute a simplification 
that uses the fewest vertices of the original curve
while approximating it according to some prescribed error criterion.
Quadratic-time algorithms have been known for some time
\cite{cc-apcmn-96} and a recent algorithm achieves time
$O(n^{4/3+\e})$ for a certain error
criterion~\cite{av-ampcu-00}.  In practice, the Douglas-Peucker
algorithm is often used as a heuristic; it can be implemented to run in time
$O(n\log n)$~\cite{hs-oidpa-94}.

\item 
How efficiently can one compute a polyhedral surface
that is an $\e$-approximation of a given triangulated surface in
$\R^3$? 
It is NP-hard to obtain the minimum-facet surface
separating two nested convex polytopes~\cite{dg-ocop3-97a},
but
polynomial-time approximation algorithms are known
(\cite{bg-aoscf-95,ms-sapo-95,as-sagp-98})
for this case, and for separating two terrain surfaces,
achieving factors within $O(1)$ or $O(\log n)$ of optimal.
However, no polynomial-time approximation results 
are known for general surfaces.

\item Given a sufficiently dense sample of
points on a surface (technically, an $\e$-sample),
reconstruct a surface homeomorphic to the original.
This has recently been accomplished for 
smooth surfaces~\cite{acdl-sahsr-00},
but remains open for surfaces 
with sharp edges and corners.

\item
Can the interior of every simply connected polyhedron whose surface is meshed
by an even number of quadrilaterals be partitioned into
a hexahedral mesh compatible with the surface meshing?~\cite{be-ecct-99}
It is known that a topological hexahedral mesh 
exists~\cite{m-cqmsw-96,e-lchmg-96}, but despite the availability of software
that extends quadrilateral surface meshes to hexahedral volume meshes,
it is not known if all hexahedral cells have planar faces.

\item Is the \emph{flip graph} connected 
for general-position points
in $\R^3$?
Given a set of $n$ points in $\R^3$,
the flip graph has a node
for each tetrahedralization of the set.
Two nodes are connected by an arc if there is a 2-to-3 or 3-to-2
``bistellar flip'' of tetrahedra between the two simplicial complexes.
In the plane, the flips correspond to convex quadrilateral diagonal switches;
in $\R^3$, a $5$-vertex convex polyhedron is ``flipped'' between two
of its tetrahedralizations.
In $\R^2$ the flip graph is connected, providing a
basis for algorithms to iterate toward the Delaunay triangulation.
A decade ago, several~\cite{epw-tpstd-90,j-ctddt-91} asked whether the
same holds in $\R^3$ (when no four points
are coplanar), but the question remains unresolved.

\item
Can every convex polytope in $\R^3$ be partitioned into tetrahedra
such that the dual graph has a Hamiltonian path?
Every convex polygon has such a \emph{Hamiltonian triangulation},
but not every nonconvex polygon does~\cite{ahms-htfr-96}.
The existence of a Hamiltonian path permits faster rendering
on a graphics screen via pipelining.

\item 
We close with
Conway's venerable thrackle conjecture, which remains open after more than
thirty years.
A \emph{thrackle} is a planar drawing of a graph of $n$ vertices
by edges
which are smooth closed curves between vertices, 
with the condition
       that any two edges intersect at exactly one point, and have distinct
       tangents there.
Conway's conjecture is that the number edges cannot exceed $n$.
Recently the upper bound was lowered from $O(n^{3/2})$ to
$2n-3$~\cite{lps-ctc-95}.
Conway offers a reward of \$1,000 for a resolution.

\end{enumerate}

\paragraph{Acknowledgement.}
We are grateful to Erik Demaine for several helpful comments.

\small

\bibliographystyle{alpha}
\bibliography{/home1/orourke/bib/geom/geom}

\newcommand{\etalchar}[1]{$^{#1}$}
\begin{thebibliography}{ECOUX95}

\bibitem[ACDL00]{acdl-sahsr-00}
N.~Amenta, S.~Choi, T.~K. Dey, and N.~Leekha.
\newblock A simple algorithm for homeomorphic surface reconstruction.
\newblock In {\em Proc. 16th Annu. ACM Sympos. Comput. Geom.}, pages 213--222,
  2000.

\bibitem[AESW91]{aesw-emstb-91}
P.~K. Agarwal, H.~Edelsbrunner, O.~Schwarzkopf, and E. Welzl.
\newblock Euclidean minimum spanning trees and bichromatic closest pairs.
\newblock {\em Discrete Comput. Geom.}, 6(5):407--422, 1991.

\bibitem[Aga92]{a-rsoas-92}
P.~K. Agarwal.
\newblock Ray shooting and other applications of spanning trees with low
  stabbing number.
\newblock {\em SIAM J. Comput.}, 21:540--570, 1992.

\bibitem[AGMR98]{agmr-vdmp-98}
G.~Albers, L.~J. Guibas, J. S.~B. Mitchell, and T.~Roos.
\newblock Voronoi diagrams of moving points.
\newblock {\em Internat. J. Comput. Geom. Appl.}, 8:365--380, 1998.

\bibitem[AGR00]{agr-lttsp-00}
N.~M. Amato, M.~T. Goodrich, and E.~A. Ramos.
\newblock Linear-time triangulation of a simple polygon made easier via
  randomization.
\newblock In {\em Proc. 16th Annu. ACM Sympos. Comput. Geom.}, pages 201--212,
  2000.

\bibitem[AH96]{ah-hgrp-96}
T.~Auer and M. Held.
\newblock Heuristics for the generation of random polygons.
\newblock In {\em Proc. 8th Canad. Conf. Comput. Geom.}, pages 38--43, 1996.

\bibitem[AHMS96]{ahms-htfr-96}
E.~M. Arkin, M. Held, J. S.~B. Mitchell, and S.~S. Skiena.
\newblock Hamiltonian triangulations for fast rendering.
\newblock {\em Visual Comput.}, 12(9):429--444, 1996.

\bibitem[Aro98]{a-ptase-98}
S. Arora.
\newblock Polynomial time approximation schemes for {Euclidean} traveling
  salesman and other geometric problems.
\newblock {\em J. ACM}, 45(5):753--782, 1998.

\bibitem[AS98]{as-sagp-98}
P.~K. Agarwal and S.~Suri.
\newblock Surface approximation and geometric partitions.
\newblock {\em SIAM J. Comput.}, 27:1016--1035, 1998.

\bibitem[AS99]{as-pckum-99}
P.~K. Agarwal and M. Sharir.
\newblock Pipes, cigars, and kreplach: The union of {M}inkowski sums in three
  dimensions.
\newblock In {\em Proc. 15th Annu. ACM Sympos. Comput. Geom.}, pages 143--153,
  1999.

\bibitem[AS00a]{as-dssga-00}
P.~K. Agarwal and M. Sharir.
\newblock {Davenport-Schinzel} sequences and their geometric applications.
\newblock In J.-R. Sack and J. Urrutia, editors, {\em
  Handbook of Computational Geometry}, pages 1--47. Elsevier Science Publishers
  B.V. North-Holland, Amsterdam, 2000.

\bibitem[AS00b]{as-pckum-00}
P.~K. Agarwal and M. Sharir.
\newblock Pipes, cigars, and kreplach: The union of {M}inkowski sums in three
  dimensions.
\newblock {\em Discrete Comput. Geom.}, 24(4):645--685, 2000.

\bibitem[AV00]{av-ampcu-00}
P.~K. Agarwal and K.~R. Varadarajan.
\newblock Approximating monotone polygonal curves using the uniform metric.
\newblock {\em Discrete Comput. Geom.}, 23, 2000.
\newblock to appear.

\bibitem[BDE{\etalchar{+}}01]{bdems-uptf-01}
M.~Bern, E.~D. Demaine, D.~Eppstein, E.~Kuo, A.~Mantler, and J.~Snoeyink.
\newblock Ununfoldable polyhedra with convex faces.
\newblock {\em Comput. Geom. Theory Appl.}, 2001.

\bibitem[BEA{\etalchar{+}}99]{be-ecct-99}
M.~Bern, D.~Eppstein, P.~K. Agarwal, N.~Amenta, P.~Chew, T.~Dey, D.~P. Dobkin,
  H.~Edelsbrunner, C.~Grimm, L.~J. Guibas, J.~Harer, J.~Hass, A.~Hicks, C.~K.
  Johnson, G.~Lerman, D.~Letscher, P.~Plassmann, E.~Sedgwick, J.~Snoeyink,
  J.~Weeks, C.~Yap, and D.~Zorin.
\newblock Emerging challenges in computational topology, 1999.
\newblock Report on an NSF Workshop on Computational Topology, June 11-12,
  Miami Beach, FL.

\bibitem[BG95]{bg-aoscf-95}
H.~Br{\"o}nnimann and M.~T. Goodrich.
\newblock Almost optimal set covers in finite {VC}-dimension.
\newblock {\em Discrete Comput. Geom.}, 14:263--279, 1995.

\bibitem[BJ00]{bj-dpcho-00}
G.~S. Brodal and R. Jacob.
\newblock Dynamic planar convex hull with optimal query time and 
$O(\log n\cdot\log\log n)$ update time.
\newblock In {\em Proc. 7th Scand. Workshop Alg. Theory}, volume
  1851 of {\em Lecture Notes in Computer Science}, pages 57--70.
  Springer-Verlag, 2000.

\bibitem[BS98]{bs-pdtcm-98}
M. Bern and A. Sahai.
\newblock Pushing disks together -- {The} continuous-motion case.
\newblock {\em Discrete Comput. Geom.}, 20:499--514, 1998.

\bibitem[BSTY98]{bsty-vdhdc-98}
J.-D. Boissonnat, M. Sharir, B. Tagansky, and M. Yvinec.
\newblock Voronoi diagrams in higher dimensions under certain polyhedral
  distance functions.
\newblock {\em Discrete Comput. Geom.}, 19(4):473--484, 1998.

\bibitem[Cap96]{c-aidp-96}
V.~Capoyleas.
\newblock On the area of the intersection of disks in the plane.
\newblock {\em Comput. Geom. Theory Appl.}, 6:393--396, 1996.

\bibitem[CC96]{cc-apcmn-96}
W.~S. Chan and F.~Chin.
\newblock Approximation of polygonal curves with minimum number of line
  segments or minimum error.
\newblock {\em Internat. J. Comput. Geom. Appl.}, 6:59--77, 1996.

\bibitem[Cha88]{c-tbsns-88}
B. Chazelle.
\newblock Tight bounds on the stabbing number of spanning trees in {Euclidean}
  space.
\newblock Report CS-TR-155-88, Dept. Comput. Sci., Princeton Univ., Princeton,
  NJ, 1988.

\bibitem[Cha91]{c-tsplt-91a}
B. Chazelle.
\newblock Triangulating a simple polygon in linear time.
\newblock {\em Discrete Comput. Geom.}, 6(5):485--524, 1991.

\bibitem[Cha96]{c-osrch-96}
T.~M. Chan.
\newblock Output-sensitive results on convex hulls, extreme points, and related
  problems.
\newblock {\em Discrete Comput. Geom.}, 16:369--387, 1996.

\bibitem[Cha97]{c-fdams-97}
B. Chazelle.
\newblock A faster deterministic algorithm for minimum spanning trees.
\newblock In {\em Proc. 38th Annu. IEEE Sympos. Found. Comput. Sci.}, page To
  appear, 1997.

\bibitem[Cha99]{c-dpcho-99}
T.~M. Chan.
\newblock Dynamic planar convex hull operations in near-logarithmic amortized
  time.
\newblock In {\em Proc. 40th Annu. IEEE Sympos. Found. Comput. Sci.}, pages
  92--99, 1999.

\bibitem[Che97]{c-nqblv-97}
L.~P. Chew.
\newblock Near-quadratic bounds for the {$L_1$} {Voronoi} diagram of moving
  points.
\newblock {\em Comput. Geom. Theory Appl.}, 7:73--80, 1997.

\bibitem[CK95]{ck-dmpsa-95}
P.~B. Callahan and S.~R. Kosaraju.
\newblock A decomposition of multidimensional point sets with applications to
  $k$-nearest-neighbors and $n$-body potential fields.
\newblock {\em J. ACM}, 42:67--90, 1995.

\bibitem[CL98]{cl-qrsam-98}
S.-W. Cheng and K.-H. Lee.
\newblock Quadtree decomposition, Steiner triangulation, and ray shooting.
\newblock In {\em {ISAAC}: 9th Internat. Sympos. Algorithms Computation}, pages
  367--376, 1998.

\bibitem[Dey98]{d-ibpks-98}
T.~K. Dey.
\newblock Improved bounds on planar $k$-sets and related problems.
\newblock {\em Discrete Comput. Geom.}, 19:373--382, 1998.

\bibitem[DG97]{dg-ocop3-97a}
G.~Das and M.~T. Goodrich.
\newblock On the complexity of optimization problems for 3-dimensional convex
  polyhedra and decision trees.
\newblock {\em Comput. Geom. Theory Appl.}, 8:123--137, 1997.

\bibitem[DMM95]{dmm-naefm-95}
M.~T. Dickerson, S.~A. McElfresh, and M.~H. Montague.
\newblock New algorithms and empirical findings on minimum weight triangulation
  heuristics.
\newblock In {\em Proc. 11th Annu. ACM Sympos. Comput. Geom.}, pages 238--247,
  1995.

\bibitem[Dye84]{d-ltatt-84}
M.~E. Dyer.
\newblock Linear time algorithms for two- and three-variable linear programs.
\newblock {\em SIAM J. Comput.}, 13:31--45, 1984.

\bibitem[ECOUX95]{eoux-ipvf-95}
V.~Estivill-Castro, J.~O'Rourke, J.~Urrutia, and D.~Xu.
\newblock Illumination of polygons with vertex floodlights.
\newblock {\em Inform. Process. Lett.}, 56:9--13, 1995.

\bibitem[Epp94]{e-amwst-94}
D.~Eppstein.
\newblock Approximating the minimum weight {Steiner} triangulation.
\newblock {\em Discrete Comput. Geom.}, 11:163--191, 1994.

\bibitem[Epp96]{e-lchmg-96}
D. Eppstein.
\newblock Linear complexity hexahedral mesh generation.
\newblock In {\em Proc. 12th Annu. ACM Sympos. Comput. Geom.}, pages 58--67,
  1996.

\bibitem[EPW90]{epw-tpstd-90}
H.~Edelsbrunner, F.~P. Preparata, and D.~B. West.
\newblock Tetrahedrizing point sets in three dimensions.
\newblock {\em J. Symbolic Comput.}, 10(3--4):335--347, 1990.

\bibitem[ES00]{es-cufop-00}
A.~Efrat and M.~Sharir.
\newblock On the complexity of the union of fat objects in the plane.
\newblock {\em Discrete Comput. Geom.}, 23:171--189, 2000.

\bibitem[GJ79]{gj-cigtn-79}
M.~R. Garey and D.~S. Johnson.
\newblock {\em Computers and Intractability: {A} Guide to the Theory of
  {NP}-Completeness}.
\newblock W. H. Freeman, New York, NY, 1979.

\bibitem[GKFS96]{gkms-lbrad-96}
D. Grigoriev, M. Karpinski, {F. M. auf der Heide}, and R.
  Smolensky.
\newblock A lower bound for randomized algebraic decision trees.
\newblock In {\em Proc. 28th ACM Sympos. Theory Comput.}, pages 612--619, 1996.

\bibitem[GO95]{go-copcg-95}
A.~Gajentaan and M.~H. Overmars.
\newblock On a class of ${O}(n^2)$ problems in computational geometry.
\newblock {\em Comput. Geom. Theory Appl.}, 5:165--185, 1995.

\bibitem[HS94]{hs-oidpa-94}
J.~Hershberger and J.~Snoeyink.
\newblock An ${O}(n \log n)$ implementation of the {Douglas}-{Peucker}
  algorithm for line simplification.
\newblock In {\em Proc. 10th Annu. ACM Sympos. Comput. Geom.}, pages 383--384,
  1994.

\bibitem[HS99]{hs-oaesp-99}
J. Hershberger and S. Suri.
\newblock An optimal algorithm for {Euclidean} shortest paths in the plane.
\newblock {\em SIAM J. Comput.}, 28(6):2215--2256, 1999.

\bibitem[Joe91]{j-ctddt-91}
B.~Joe.
\newblock Construction of three-dimensional {Delaunay} triangulations using
  local transformations.
\newblock {\em Comput. Aided Geom. Design}, 8(2):123--142, May 1991.

\bibitem[Kar84]{k-nptal-84a}
N.~Karmarkar.
\newblock A new polynomial-time algorithm for linear programming.
\newblock {\em Combinatorica}, 4:373--395, 1984.

\bibitem[Kha80]{k-palp-80}
L.~G. Khachiyan.
\newblock Polynomial algorithm in linear programming.
\newblock {\em U.S.S.R. Comput. Math. and Math. Phys.}, 20:53--72, 1980.

\bibitem[KMM97]{kmm-eaesp-97}
S.~Kapoor, S.~N. Maheshwari, and J. S.~B. Mitchell.
\newblock An efficient algorithm for {Euclidean} shortest paths among polygonal
  obstacles in the plane.
\newblock {\em Discrete Comput. Geom.}, 18:377--383, 1997.

\bibitem[Kol01]{k-atubv-01}
V. Koltun.
\newblock Almost tight upper bounds for vertical decompositions in four
  dimensions.
\newblock In {\em Proc. 42nd Sympos. Foundations Comput. Sci.}, 2001.

\bibitem[LK96]{lk-qgtam-96}
C. Levcopoulos and D. Krznaric.
\newblock Quasi-greedy triangulations approximating the minimum weight
  triangulation.
\newblock In {\em Proc. 7th ACM-SIAM Sympos. Discrete Algorithms}, pages
  392--401, 1996.

\bibitem[LPS95]{lps-ctc-95}
L. Lov{\'a}sz, J. Pach, and M. Szegedy.
\newblock On Conway's thrackle conjecture.
\newblock In {\em Proc. 11th Annu. ACM Sympos. Comput. Geom.}, pages 147--151,
  1995.

\bibitem[Meg84]{m-lpltw-84}
N.~Megiddo.
\newblock Linear programming in linear time when the dimension is fixed.
\newblock {\em J. ACM}, 31:114--127, 1984.

\bibitem[Mit96]{m-cqmsw-96}
S.~A. Mitchell.
\newblock A characterization of the quadrilateral meshes of a surface which
  admit a compatible hexahedral mesh of the enclosed volume.
\newblock In {\em Proc. 13th Sympos. Theoret. Aspects Comput. Sci.}, volume
  1046 of {\em Lecture Notes Comput. Sci.}, pages 465--476. Springer-Verlag,
  1996.

\bibitem[Mit00]{m-gspno-00}
J. S.~B. Mitchell.
\newblock Geometric shortest paths and network optimization.
\newblock In J.-R. Sack and J. Urrutia, editors, {\em
  Handbook of Computational Geometry}, pages 633--701. Elsevier Science
  Publishers B.V. North-Holland, Amsterdam, 2000.

\bibitem[MRW92]{mrw-mlpop-92}
J. S.~B. Mitchell, G. Rote, and G.~Woeginger.
\newblock Minimum-link paths among obstacles in the plane.
\newblock {\em Algorithmica}, 8:431--459, 1992.

\bibitem[MS95]{ms-sapo-95}
J. S.~B. Mitchell and S. Suri.
\newblock Separation and approximation of polyhedral objects.
\newblock {\em Comput. Geom. Theory Appl.}, 5:95--114, 1995.

\bibitem[MSW96]{msw-sblp-96}
J.~Matou{\v s}ek, M. Sharir, and E. Welzl.
\newblock A subexponential bound for linear programming.
\newblock {\em Algorithmica}, 16:498--516, 1996.

\bibitem[O'R93]{o-cgc18-93}
J.~O'Rourke.
\newblock Computational geometry column 18.
\newblock {\em Internat. J. Comput. Geom. Appl.}, 3(1):107--113, 1993.
\newblock Also in {\em SIGACT News\/} 24:1 (1993), 20--25.

\bibitem[O'R00]{o-fucg-00}
J. O'Rourke.
\newblock Folding and unfolding in computational geometry.
\newblock In {\em Proc. Japan Conf. Discrete Comput. Geom. '98}, volume 1763 of
  {\em Lecture Notes Comput. Sci.}, pages 258--266. Springer-Verlag, 2000.

\bibitem[OS97]{os-vepvgcr-97}
J.~O'Rourke and I.~Streinu.
\newblock Vertex-edge pseudo-visibility graphs: Characterization and
  recognition.
\newblock In {\em Proc. 13th Annu. ACM Sympos. Comput. Geom.}, pages 119--128,
  1997.

\bibitem[PSS01]{pss-ucc3d-01}
J. Pach, I. Safruit, and M. Sharir.
\newblock The union of congruent cubes in three dimensions.
\newblock In {\em Proc. 17th ACM Sympos. Comput. Geom.}, pages 19--28, 2001.

\bibitem[PY90]{py-ebsph-90}
M.~S. Paterson and F.~F. Yao.
\newblock Efficient binary space partitions for hidden-surface removal and
  solid modeling.
\newblock {\em Discrete Comput. Geom.}, 5:485--503, 1990.

\bibitem[SA95]{sa-dsstg-95}
M. Sharir and P.~K. Agarwal.
\newblock {\em {Davenport}-{Schinzel} Sequences and Their Geometric
  Applications}.
\newblock Cambridge University Press, New York, 1995.

\bibitem[Sei91]{s-sfira-91}
R.~Seidel.
\newblock A simple and fast incremental randomized algorithm for computing
  trapezoidal decompositions and for triangulating polygons.
\newblock {\em Comput. Geom. Theory Appl.}, 1(1):51--64, 1991.

\bibitem[Sha94]{s-atubl-94}
M. Sharir.
\newblock Almost tight upper bounds for lower envelopes in higher dimensions.
\newblock {\em Discrete Comput. Geom.}, 12:327--345, 1994.

\bibitem[She75]{s-cpcn-75}
G.~C. Shephard.
\newblock Convex polytopes with convex nets.
\newblock {\em Math. Proc. Camb. Phil. Soc.}, 78:389--403, 1975.

\bibitem[Sno97]{s-pl-97}
J.~Snoeyink.
\newblock Point location.
\newblock In J.~E. Goodman and J. O'Rourke, editors, {\em Handbook of
  Discrete and Computational Geometry}, chapter~30, pages 559--574. CRC Press
  LLC, Boca Raton, FL, 1997.

\bibitem[T\'ot00]{t-psmks-00}
G.~T\'oth.
\newblock Point sets with many $k$-sets.
\newblock In {\em Proc. 16th Annu. ACM Sympos. Comput. Geom.}, pages 37--42,
  2000.

\bibitem[T\'ot01]{t-nbpp-01}
C.~D. T\'oth.
\newblock A note on binary plane partitions.
\newblock In {\em Proc. 17th ACM Sympos. Comput. Geom.}, pages 151--156, 2001.

\bibitem[Urr00]{u-agip-00}
J. Urrutia.
\newblock Art gallery and illumination problems.
\newblock In J.-R. Sack and J. Urrutia, editors, {\em
  Handbook of Computational Geometry}, pages 973--1027. North-Holland, 2000.

\bibitem[Vai89]{v-ghm-89}
P.~M. Vaidya.
\newblock Geometry helps in matching.
\newblock {\em SIAM J. Comput.}, 18:1201--1225, 1989.

\bibitem[Wel93]{w-ggssn-93}
E. Welzl.
\newblock Geometric graphs with small stabbing numbers: {Combinatorics} and
  applications.
\newblock In {\em Proc. 9th Internat. Conf. Fund. Comput. Theory}, Lecture
  Notes Comput. Sci., Springer-Verlag, 1993.

\bibitem[ZSSM96]{zssm-grpgv-96}
C.~Zhu, G.~Sundaram, J.~Snoeyink, and J. S.~B. Mitchell.
\newblock Generating random polygons with given vertices.
\newblock {\em Comput. Geom. Theory Appl.}, 6:277--290, 1996.

\end{thebibliography}

\end{document}